\documentclass[11pt,twoside]{article}

\usepackage{asp2006}
\usepackage{epsf}
\usepackage{psfig}
\usepackage{lscape}

\begin{document}
\title{Unveiling Galaxy Interactions: Watching the Tides Roll}   %%% Fill in title
\author{William C. Keel}   %%% Fill in author names
\affil{University of Alabama}    %%% Fill in author affiliations

\begin{abstract} %%% Abstract to run on from here.
I set the stage for discussion of the stellar populations in interacting galaxies by looking back over the slow development of our understanding of these systems. From early anecdotal collections, to systematic cataloging, and finally to increasingly sophisticated $n$-body calculations, we have seen how gravity in distributed systems can produce the stunning variety of structures we see. At the same time, measures across the spectrum have made it clear that galaxy interactions are linked to star formation, albeit with the physical mechanisms much less clear. Improved data sets, including HST imaging, deep IR data, and large samples with well-defined selection criteria, have started to reveal correlations with dynamical parameters pointing to detailed histories of starbirth during collisions. The merger hypothesis for elliptical galaxies has broadened into seeing interactions and mergers as important parts of the overall evolution of galaxies. The connection becomes more important as we look to higher redshift, where more frequent interactions can drive the evolution of galaxies in multiple ways. Links between the properties of central black holes and surrounding galaxies makes it important likewise to understand the connections between AGN and interactions, which has remained more ambiguous due to the strong role of sample selection. Finally, contemporary data reach deep enough to show that most galaxies have interacted in the observable past; we must consider these events to be a normal part of galaxy history.
\end{abstract}

%%% MAIN BODY OF TEXT GOES HERE. CONSULT "INSTRUCTIONS FOR AUTHORS USING
%%% LATEX2E MARKUP", SECTIONS 2.3-2.6 FOR HELP WITH EQUATIONS, FIGURES,
%%% AND TABLES.

%\section{}   %%% Top level section head (remove "%" symbol)
%\subsection{}   %%% Second level section head (remove "%" symbol)
%\subsubsection{}   %%% Lowest level section head (remove "%" symbol)
%\section*{}    %%% Unnumbered top level section head (remove "%" symbol)
%\subsection*{}   %%% Unnumbered second level section head (remove "%" symbol)

\section{Galaxies interact}
With hindsight, astronomers knew that galaxies interact before understanding what galaxies are. The famous depiction of the Whirlpool Galaxy M51 and its companion by the Earl of Rosse, from 1845, marks not only the clear discovery of spiral structure in galaxies, but indicates a connection of the spiral pattern to the companion object. Within 20 years, Foucault's improvements (notably silver coating of glass mirrors) allowed Jean Chacornac to depict considerably more detailed structure, including the dust patches seen against the companion, with the much smaller Paris 0.8m reflector (\citealp{TH2008}).

Many astronomers' attention was drawn to the remarkable variety of forms taken by interacting galaxies on viewing the atlases produced by \citet{VV59}, with its extension by \citet{VV77}, and especially the \citet{Arp1966} atlas. Vorontsov-Velyaminov's work was somewhat neglected in the West, owing at least partly to its publication in a very small photographically-reproduced edition with images reproduced from the Palomar Sky Survey. In contrast, Arp's images, many taken with the Palomar 5m telescope and using considerable care in acquisition and reproduction, were especially striking. Their arrangement in the atlas emphasized bridges, tails, distorted spiral patterns, and other features not well represented in such collections as the {\it Hubble Atlas} \citep{Sandage1961}. While these atlases made it obvious that galaxy encounters can affect their structures profoundly, it was not so obvious that gravity alone was at work here. Particularly with some of the very long tails and bridges (as in NGC 4038/9, the Antennae, or the bridging filament spanning over 100 kpc between components of Arp 295 and then beyond), there was serious discussion of whether magnetic fields or more exotic processes might be shaping the observed structures.

The growing evidence for dark matter on galaxies, and the clear value of statistically well-defined sample of binary galaxies as probes of their masses to large radii, helped drive the creation of uniform catalogs of paired galaxies, whether showing signs of interactions or not. The Catalog of Isolated Pairs of Galaxies by \citet{kpairs} brought out several issues which continue to be important in sample construction - contamination by group members seen at fortuitous angles, how to distinguish distorted single and double galaxies, the importance of well-defined isolation criteria for pairs, and the incidence of redshift interlopers. Completion of redshift measurements for the 585 clear pairs in this sample spurred work on counterparts for the southern hemisphere (e.g., \citealp{RR1995}. Large-scale redshift surveys have allowed incorporation of more complete isolation information for pairs (e.g., \citealp{Barton2000}).

\section{Tides bring about distortions and mergers}

Erik Holmberg's recognition that galaxies form multiple systems much more often then seen in a random distribution led him to consider whether dynamical processes could allow capture of galaxies. To this end, he conducted a series of analog computations, substituting the inverse-square behavior of light for gravity \citep{Holmberg1941} (also see the appreciation by \citealp{Rood1987}). These results made it clear that sufficient energy can be exchanged between the galaxies' orbits and internal motions to allow initially parabolic encounters to yield bound systems. They also provided the first hints of how tidal encounters can produce distortions of spiral structure, although this effect was at the limit of what could be interpreted from a limited number of test particles. These issues were revisited numerically in the seminal work of \citet{TT1972}, who used restricted three-body techniques not only to show how tides in rotating disks could yield spectacular disruptions, but to model the specific distortions seen in M51, the Antennae NGC 4038/9, and the Mice NGC 4676.

Broad energy considerations from these and later simulations, the late-time structure of simulated encounter products, and some simple statistics of merger signatures among NGC galaxies led \citet{Toomre1977} to state what came to be known as the ``merger hypothesis". Since the remnants of major mergers between disk galaxies often end up structurally very much like elliptical galaxies, does this suggest that many (or perhaps all) elliptical galaxies have such an origin? Toomre even arranged well-known NGC systems into an order known to proposal reviewers the world over as the Toomre sequence, providing snapshots of various stages from initial encounter to final wrapping of tidal tails into the remnant. 

As the zoo of interactions has been explored, important subsets have emerged. Ring galaxies can be the kinematic aftermath of deep plunging encounters when dense companions pass through the disks of spirals. Polar rings may be the debris of tidal distortion which has settled into slowly-precessing paths near the disk poles. Shells of stars in the halos of early-type galaxies are well-modelled as the remnant of low-mass, dynamically cold systems after minor mergers. Asymmetries and tail, of course, can trace strong interactions; there has been recent work in applying quantitative methods to their interpretations (such as the CAS system, \citealp{Conselice}). All these signatures may be more pronounced, especially for weak interactions, in the H I structure, which is often very extended and correspondingly fragile, but wholesale understanding of H I as a tidal tracer is only beginning \citep{Hibbard}.

\section{Interactions drive star formation}
Anecdotal evidence had suggested for some time interacting and merging galaxies changed more than their shapes. The prototypical starburst galaxies M82 and NGC 7714 are in strongly interacting systems, and suspicious numbers of Markarian galaxies (whether dominated by star formation or AGN) likewise show peculiar morphology. Direct connection to star-forming histories was made by \citet{LT1978}, who noted that the scatter in {\it UBV} colors of interacting galaxies from the Arp atlas was larger than for normal galaxies from the {\it Hubble Atlas}, and that this scatter could be explained neatly by ongoing or fading bursts of star formation (presumably triggered by the interaction). In retrospect, this fit with the Holmberg effect (noted by \citealp{Holmberg1958}), a tendency for the colors of paired galaxies to correlate with each other more strongly than their morphologies would require (distinct from the other Holmberg effect, an overdensity of satellite galaxies along their primaries' projected minor axes).

Throughout the 1980s, several groups compared tracers of star formation (H$\alpha$, near-UV excess, far-infrared emission, radio continuum) between samples of interacting and noninteracting galaxies (\citealp{KK1984}, \citealp{KKHH}, \citealp{Bushouse1986}, \citealp{KRKHH}, \citealp{Bushouse1987}, \citealp{HHKK}). The robust conclusion was that star-formation rates are higher in interacting galaxies than in otherwise similar galaxies in sparse environments. This applies to both nuclei and to galaxy disks as a whole.
Sample selection has remained important in such comparisons, being advanced considerably by large-scale redshift surveys (\citealp{Barton2000}, \citealp{Barton2003}). These samples have shown hints that the star formation lags the strongest tidal perturbation. Application of halo formalism \citep{Barton2007} has shed light on the differences among various samples of paired galaxies, since they tend to be in group environments which may suppress star formation compared to truly isolated galaxies.

With the all-sky opening of the far-infrared window by IRAS starting in late 1983, the prevalence of the starburst phenomenon became apparent. The class of ultraluminous infrared galaxies (ULIRGs, among other acronyms) could be identified, reaching to bolometric luminosities above 10$^{12}$ solar luminosities with most emerging only in the far-infrared. Determining their energy source (essentially, AGN or very massive starbursts) was hampered by the strong extinction, so that the usual optical diagnostics might not be definitive. These objects occur preferentially in merging or strongly interacting systems, at least at the present epoch. This uniquely associates the most powerful recent star-forming events with a subset of galaxy interactions, often considered to be the final inspiral during mergers of massive gas-rich disks. 
The dominant energy source in many of these objects has remained ambiguous even in the era of Spitzer spectroscopy  \citep{Veilleux2009}. It is certainly interesting that so many have comparable energy input from AGN and starburst components.

\section{Star clusters trace the history of starbirth}

A nagging problem with the ``merger hypothesis" had long been posed by the statistics of globular clusters. Elliptical galaxies typically have a specific frequency (number of globulars normalized to $B$-band luminosity) too high to make them by adding the contents of spiral galaxies, unless mergers not only drive starbursts but the stars in these bursts are preferentially formed in very massive and long-lived clusters. That is, the hypothesis works only if galaxy mergers make globular clusters more efficiently than individual massive stars. This was one reason for the great interest attracted by the discovery of rich populations of very luminous young star clusters (super star clusters, SSCs), which are particularly common in interacting and merging systems. Ground-based data had shown that some interacting systems contain very luminous star-forming regions, but not that many of these are single compact clusters rather than more widespread  associations or complexes (in the Local group, the distinction between NGC 604 in M33 and 30 Doradus in the LMC) This was above all a product of data from the {\it Hubble Space Telescope} (HST), although there had been hints of the nature of a few of these objects from ground-based work. The most extensive results has been on the cluster system in the Antennae (NGC 4038/9, \citealp{Whitmore1995}, \citealp{Zhang}).

HST data can provide inventories of bright clusters for systems to nearly 100 Mpc. With extensive modeling of cluster evolution in spectrum and luminosity, color-magnitude diagrams of cluster populations have taken on some of the role played by the Hertzsprung-Russell diagram in unraveling the history and population of stars. These populations have been used to estimate burst ages and test models of cluster disruption, an important issue in learning whether the power-law luminosity function of young clusters may evolve to the Gaussian form characteristic of old globular clusters.

\section{What drives the star formation?}
There has been no shortage of suggestions for what physical mechanisms might enhance star formation during galaxy encounters. A selection of these follows. Trying to discriminate among their potential roles has proven challenging, since many would normally act in the same systems. The available clues include the behavior of star formation in galaxies with various kinds of velocity structure and in direct versus retrograde encounters \citep{Keel1993}, behavior of star formation as functions of projected separation and velocity \citep{Barton2003}, and reconstruction of individual interactions using simulations constrained by morphological and kinematic data (an approach dating broadly to \citealp{TT1972}, and set out recently in the ``Identikit" approach of \citealp{Identikit}). 

\subsection{Cloud collisions}

It seems eminently plausible that star formation can be triggered by collisions between molecular clouds. The rate of such collisions might be increased dramatically by even minor tidal effects, since orbit crossings of clouds on initially similar orbits would suffice. Since there is evidence for enhanced star formation in galaxy pairs which are too far apart to have undergone physical collisions, this wold be an attractive way to alter internal processes. Such collisions within clouds of a single disk might be most effective (most frequent and at highest velocity) when the disk has been disturbed and warped so as to lead to orbit crossing by clouds over a large region, most often an arc such as may be seen in NGC 5257 and 6745.

\subsection{ISM phase change/pressure driving}
The interstellar medium in galaxies is a complex multi-phase environment, where changes in one phase can alter conditions in the others. \citet{JS1992} considered this interplay in the context of interactions, noting that a pressure change in the hot ``intercloud" medium could propagate rapidly through a galaxy (faster than the dynamical timescale associated with most other proposed agents). The star formation would then result from a change in the pressure balance between dense clouds and the surrounding hot material.

\subsection{Induced gas flows}
Bars are well-known to channel gas radially, transferring angular momentum outward and providing one avenue for feeding material inward by roughly an order of magnitude in radius. Simulations with appropriate initial conditions (\citealp{Noguchi1988}, \citealp{BH1996}) show that some encounters can create not only long-lasting stellar bars, but transient, gas-rich bars which channel gas inward, providing ample material for star formation, and leaving little trace in the stellar distribution. The latter point is crucial, since near-IR surveys have not uncovered large numbers of obscured bars in interacting systems.

\subsection{Toomre-style disk instabilities}
\citet{Toomre1964} derived a stability criterion for a self-gravitating disk of particles, which has later been applied to purely gaseous as well as stellar disks. This has been invoked to explain the cutoff in star formation seen in normal disks, when the disk changes from stable to unstable (\citealp{RCK1989}, anticipated from a somewhat different formulation by \citealp{ZS1988}). The relevance to interactions comes from the possibility that kinematic disturbance could change the local surface density or velocity dispersion so as to render large amounts of gas unstable in short times. An example may be found in NGC 6621/2 \citep{KB2003}, in which the interaction has reversed the slope of the velocity curve across a region of the strongest star formation in the whole system.

\subsection{Cross-fuelling}
Dumping of gas between members of a galaxy pair could provide early-type galaxies with fresh reservoirs of material for star formation, and need not face the same angular-momentum barrier as radial motion within a single disk. Its effects would be most clear in mixed-morphology pairs, where large amounts of gas in the E/S0 galaxy could be assumed to have come from the spiral galaxy. There are a few examples of pairs in which mass transfer is evidently seen, but signs that the material either forms stars or fuels AGN remain weak (\citealp{WCK2004}, \citealp{DSA2005}).

\section{Active galactic nuclei, interactions, and starbursts}
In contrast to the strength of the statistical evidence linking galaxy interactions to star formation, the analogous question for active galactic nuclei remains unresolved. Why bring this up at a meeting dealing specifically with star formation in these systems? These two issues must be connected. The relations between stellar mass and velocity dispersion of galaxy spheroids and masses of central black holes (\citealp{Magorrian}, \citealp{Gebhardt}) indicate that whatever the growth history of spheroids, the black holes must roughly track them.
Major episodes of star formation should then be accompanied (not necessarily at quite the same time) by accretion episodes for the black hole. At the highest luminosities, the comparable space densities of ULIRGs and quasars led \citet{Sanders1988} to speculate that they might be related in an evolutionary sequence. This scheme has continued to make sense with a broad range of later data. Broadly, major mergers of gas-rich disks create massive kpc-scale reservoirs of dense gas, hosting correspondingly massive and compact starbursts. The mass loss from these stars adds to the available material to feed the central object, whose accretion luminosity may start to rise during the starburst phase. Early on, it will be obscured in the optical and near-IR, but feedback will begin clearing its surroundings, so that typical QSOs are seen later in the postmerger phases as the dust dissipates.

The observational connection between interacting systems and AGN has been reported by various studies to be strong, weak, or nonexistent. Perhaps even more than in studying star formation, details of sample selection prove to be crucial. \citet{KK1984} found an excess of Seyfert nuclei in a sample of interacting galaxies, as did \citet{Dahari} and \citet{Bushouse1986}; these studies found hints that AGN were most common in galaxies that are close in projection but not (yet) highly distorted. Incorporating an attempt to match a control sample by Hubble type and luminosity, \citet{FWS1988} found that an excess of companions to Seyfert galaxies was confined to low-luminosity companions. Perhaps because details of sample selection can be at once crucial and subtle, subsequent studies have failed to yield a consensus on the relation between AGN and interactions. \citet{LS1995} find that any excess of companion galaxies is confined to type 2 Seyferts. Powerful radio galaxies show signs of past interactions (boxy isophotes, shells, tails) more often than ellipticals of the same optical luminosity \citep{TMH1986}, with hints that the connection is stronger for FR II sources \citep{Zirbel}. Visual classifications of over 3000 merging systems from the SDSS suggest no excess of AGN in this sample \citep{Darg}.

At higher luminosities, early HST imaging of QSO host galaxies at $z<0.5$ showed a large incidence of companion galaxies and tidal features \citep{Bahcall97}.
Some elliptical QSO hosts show shell structure \citep{Bennert}, evidence of minor mergers. Spectroscopy of QSO host galaxies often shows evidence of recent star formation \citep{Canalizo}, consistent with triggering by the same tidal events (although evidence on the timing between starburst and QSO phases remains scant). These factors fit with the Sanders et al. conjecture relating merging, starbursts, and feeding of an AGN; even so, there is much to be done on the physical mechanisms, extension to minor interactions, and cosmic history of these events.

An important development has been the result of comparison of H I morphologies between Seyfert and inactive galaxies by \citet{Kuo2008}, showing that at least 2/3 of the Seyferts they examined have H I distortions traceable to tidal interactions, compared to only 15\% of a matched control sample, and that the H I structure is disturbed much more often than the optical morphology.

\section{Interactions everywhere, all the time?}
We need to free our heads of the idea that galaxies are normally equilibrium systems, disturbed only rarely as an aberration. We can now look deeply enough that most galaxies reveal a history of interactions. In the Local Group, we see the Magellanic Stream, and recently-detected tails of stars from the Andromeda galaxy and M33 \citep{Ibata2007}. Deep imaging of cluster environments, with careful suppression of stray light, reveals tails of stars writing billions of years' worth of interaction history (e.g., \citealp{Mihos}) and eventually becoming the diffuse intracluster starlight comprising 10--25\% of the stars in some clusters \citep{Feldmeier}. Interactions are ubiquitous parts of the lives of galaxies.

\acknowledgements I thank the meeting organizers for the opportunity to present this review. Attendance was supported by a UA Dean's Leadership Board Faculty Fellowship.

\end{document}